\documentclass[twocolumn,showpacs,aps,prl,superscriptaddress]{revtex4}

\usepackage{graphicx}
\usepackage{dcolumn}
\usepackage{epsfig}
\usepackage{amsmath}

\newcommand{\BaBarYear}    {04}
\newcommand{\BaBarNumber}  {020}

\newcommand{\SLACPubNumber} {10542}
\newcommand{\LANLNumber} {0407013}

 \newcommand{\BaBarType}      {PUB}  

\input pubboard/babarsym
\RequirePackage{xspace}

\newcommand{\pvec}{{\bf p}}

\newcommand{\calB}{\ensuremath{{\cal B}}}

\newcommand{\bfemsix}{${\cal B}(10^{-6}$)}

\newcommand{\DE}{\ensuremath{\Delta E}}

\newcommand{\xf}{\ensuremath{{\cal F}}}

\newcommand{\thetaT}{\ensuremath{\theta_{\rm T}}}
\newcommand{\costhr}{\ensuremath{\cos\thetaT}}

\newcommand\etal{{\it et al.}}
\newcommand{\half}{\ensuremath{{1\over2}}}

\newcommand{\bma}[1]{\boldmath{$#1$}}

\newcommand{\bfig}{\begin{figure}[htbpc!]}
\newcommand{\efig}{\end{figure}}
\newcommand\bef{\begin{figure}}
\newcommand\edf{\end{figure}}
\newcommand\dbline{\noalign{\vskip 0.10truecm\hrule}\noalign{\vskip 2pt}\noalign{\hrule\vskip 0.10truecm}}
\providecommand{\tbline}{\noalign{\vskip 0.05truecm\hrule\vskip0.05truecm}}

\newcommand\beq{\begin{equation}}
\newcommand\eeq{\end{equation}}
\newcommand\bear{\begin{array}}
\newcommand\enar{\end{array}}
\newcommand\beqa{\begin{eqnarray}}
\newcommand\eeqa{\end{eqnarray}}
\newcommand\ben{\begin{enumerate}}
\newcommand\een{\end{enumerate}}

\newcommand{\UfourS}{\ensuremath{\Upsilon(4S)}}

\newcommand{\etagg}{\ensuremath{\eta_{\gaga}}}
\newcommand{\etappp}{\ensuremath{\eta_{3\pi}}}
\newcommand{\etatogg}{\ensuremath{\eta\ra\gaga}}
\newcommand{\etatoppp}{\ensuremath{\eta\ra\pi^+\pi^-\pi^0}}

\newcommand{\Kst}{\ensuremath{K^*}}

\newcommand{\kzs}{\ensuremath{\KS}}

\newcommand{\kstopipi}{\ensuremath{\KS\to\pi^+\pi^-}}

\newcommand{\hp}{\ensuremath{h^{+}}}

\newcommand{\azero}{\ensuremath{a_0}}

\newcommand{\azeroz}{\ensuremath{a_0^0}}
\newcommand{\fazeroK}{\ensuremath{a_0 K}}
\newcommand{\fazeroKz}{\ensuremath{a_0 K^0}}
\newcommand{\fazeropi}{\ensuremath{a_0\pi}}

\newcommand{\fazeromKz}{\ensuremath{a_0^-\Kzb}}
\newcommand{\fazerozKz}{\ensuremath{a_0^0\Kz}}

\newcommand{\fazeromhp}{\ensuremath{a_0^- h^+}}
\newcommand{\fazerozhp}{\ensuremath{a_0^0 h^+}}
\newcommand{\fazeromKp}{\ensuremath{a_0^- K^+}}
\newcommand{\fazerozKp}{\ensuremath{a_0^0 K^+}}
\newcommand{\fazerompip}{\ensuremath{a_0^-\pi^+}}
\newcommand{\fazerozpip}{\ensuremath{a_0^0\pi^+}}

\newcommand{\azeromKz}{\ensuremath{\Bm\ra\fazeromKz}}
\newcommand{\azerozKz}{\ensuremath{\Bz\ra\fazerozKz}}
\newcommand{\azeroKz}{\ensuremath{\B\ra\fazeroKz}}

\newcommand{\azeroK}{\ensuremath{\B\ra\fazeroK}}
\newcommand{\azeropi}{\ensuremath{\B\ra\fazeropi}}
\newcommand{\azeromKp}{\ensuremath{\Bz\ra\fazeromKp}}
\newcommand{\azerozKp}{\ensuremath{\Bp\ra\fazerozKp}}
\newcommand{\azerompip}{\ensuremath{\Bz\ra\fazerompip}}
\newcommand{\azerozpip}{\ensuremath{\Bp\ra\fazerozpip}}

\newcommand{\fazeromksgg}{\ensuremath{a_0^-(\etagg)\Kzb}}
\newcommand{\fazeromksthrpi}{\ensuremath{a_0^-(\etappp)\Kzb}}

\newcommand{\fazerozksgg}{\ensuremath{a_0^0(\etagg)K^0}}
\newcommand{\fazerozksthrpi}{\ensuremath{a_0^0(\etappp)K^0}}

\newcommand{\fazeromhpgg}{\ensuremath{a_0^-(\etagg)\hp}}
\newcommand{\fazeromKpgg}{\ensuremath{a_0^-(\etagg)\Kp}}
\newcommand{\fazerompipgg}{\ensuremath{a_0^-(\etagg)\pip}}
\newcommand{\fazeromhpthrpi}{\ensuremath{a_0^-(\etappp)\hp}}
\newcommand{\fazeromKpthrpi}{\ensuremath{a_0^-(\etappp)\Kp}}
\newcommand{\fazerompipthrpi}{\ensuremath{a_0^-(\etappp)\pip}}

\newcommand{\fazerozhpgg}{\ensuremath{a_0^0(\etagg)\hp}}
\newcommand{\fazerozKpgg}{\ensuremath{a_0^0(\etagg)\Kp}}
\newcommand{\fazerozpipgg}{\ensuremath{a_0^0(\etagg)\pip}}
\newcommand{\fazerozhpthrpi}{\ensuremath{a_0^0(\etappp)\hp}}
\newcommand{\fazerozKpthrpi}{\ensuremath{a_0^0(\etappp)\Kp}}
\newcommand{\fazerozpipthrpi}{\ensuremath{a_0^0(\etappp)\pip}}

\newcommand{\BazeromKz}{\ensuremath{\calB(\azeromKz)}}
\newcommand{\razeromKz}{\ensuremath{xx^{+xx}_{-xx}\pm xx}}

\newcommand{\ulazeromKz}{\ensuremath{xx}\xspace}
\newcommand{\ULazeromKz}{\ensuremath{\ulazeromKz\times 10^{-6}}\xspace}
\newcommand{\sazeromKz}{\ensuremath{xx}\xspace}

\newcommand{\BazerozKz}{\ensuremath{\calB(\azerozKz)}}
\newcommand{\razerozKz}{\ensuremath{xx^{+xx}_{-xx}\pm xx}}

\newcommand{\ulazerozKz}{\ensuremath{xx}\xspace}
\newcommand{\ULazerozKz}{\ensuremath{\ulazerozKz\times 10^{-6}}\xspace}
\newcommand{\sazerozKz}{\ensuremath{xx}\xspace}

\newcommand{\BazeromKp}{\ensuremath{\calB(\azeromKp)}}
\newcommand{\razeromKp}{\ensuremath{xx^{+xx}_{-xx}\pm xx}}

\newcommand{\ulazeromKp}{\ensuremath{xx}\xspace}
\newcommand{\ULazeromKp}{\ensuremath{\ulazeromKp\times 10^{-6}}\xspace}
\newcommand{\sazeromKp}{\ensuremath{xx}\xspace}

\newcommand{\BazerozKp}{\ensuremath{\calB(\azerozKp)}}
\newcommand{\razerozKp}{\ensuremath{xx^{+xx}_{-xx}\pm xx}}

\newcommand{\ulazerozKp}{\ensuremath{xx}\xspace}
\newcommand{\ULazerozKp}{\ensuremath{\ulazerozKp\times 10^{-6}}\xspace}
\newcommand{\sazerozKp}{\ensuremath{xx}\xspace}

\newcommand{\Bazerompip}{\ensuremath{\calB(\azerompip)}}
\newcommand{\razerompip}{\ensuremath{xx^{+xx}_{-xx}\pm xx}}

\newcommand{\ulazerompip}{\ensuremath{xx}\xspace}
\newcommand{\ULazerompip}{\ensuremath{\ulazerompip\times 10^{-6}}\xspace}
\newcommand{\sazerompip}{\ensuremath{xx}\xspace}

\newcommand{\Bazerozpip}{\ensuremath{\calB(\azerozpip)}}
\newcommand{\razerozpip}{\ensuremath{xx^{+xx}_{-xx}\pm xx}}

\newcommand{\ulazerozpip}{\ensuremath{xx}\xspace}
\newcommand{\ULazerozpip}{\ensuremath{\ulazerozpip\times 10^{-6}}\xspace}
\newcommand{\sazerozpip}{\ensuremath{xx}\xspace}

\providecommand{\bfemsix}{${\cal B} (10^{-6})$}

\renewcommand{\razerompip}{\ensuremath{2.8^{+1.5}_{-1.3}\pm 0.7}}
\renewcommand{\ulazerompip}{\ensuremath{5.1}\xspace}
\renewcommand{\sazerompip}{\ensuremath{2.0}\xspace}

\renewcommand{\razeromKp}{\ensuremath{0.4^{+1.0}_{-0.8}\pm 0.2}}
\renewcommand{\ulazeromKp}{\ensuremath{2.1}\xspace}
\renewcommand{\sazeromKp}{\ensuremath{0.4}\xspace}

\renewcommand{\razeromKz}{\ensuremath{-1.5^{+2.4}_{-1.8}\pm 0.8}}
\renewcommand{\ulazeromKz}{\ensuremath{3.9}\xspace}
\renewcommand{\sazeromKz}{\ensuremath{0.6}\xspace}

\renewcommand{\razerozpip}{\ensuremath{2.6^{+2.0}_{-1.7}\pm 1.0}}
\renewcommand{\ulazerozpip}{\ensuremath{5.8}\xspace}
\renewcommand{\sazerozpip}{\ensuremath{1.4}\xspace}

\renewcommand{\razerozKp}{\ensuremath{0.4^{+1.1}_{-0.7}\pm 0.3}}
\renewcommand{\ulazerozKp}{\ensuremath{2.5}\xspace}
\renewcommand{\sazerozKp}{\ensuremath{0.4}\xspace}

\renewcommand{\razerozKz}{\ensuremath{2.8^{+3.1}_{-2.4}\pm 1.1}}
\renewcommand{\ulazerozKz}{\ensuremath{7.8}\xspace}
\renewcommand{\sazerozKz}{\ensuremath{1.0}\xspace}

\begin{document}

\preprint{\babar-PUB-\BaBarYear/\BaBarNumber} 
\preprint{SLAC-PUB-\SLACPubNumber} 

\begin{flushleft}
\babar-\BaBarType-\BaBarYear/\BaBarNumber \\
SLAC-PUB-\SLACPubNumber\\
hep-ex/\LANLNumber
\end{flushleft}

\par\vskip .2cm

\title{
 \large \bf\boldmath Search for $B$-Meson Decays to Two-body Final States with \azero(980) Mesons
}

%
\author{B.~Aubert}
\author{R.~Barate}
\author{D.~Boutigny}
\author{F.~Couderc}
\author{J.-M.~Gaillard}
\author{A.~Hicheur}
\author{Y.~Karyotakis}
\author{J.~P.~Lees}
\author{V.~Tisserand}
\author{A.~Zghiche}
\affiliation{Laboratoire de Physique des Particules, F-74941 Annecy-le-Vieux, France }
% ...
\collaboration{The \babar\ Collaboration}
\noaffiliation

\date{\today}

\begin{abstract}
We present a search for $B$ decays to charmless final 
states involving charged or neutral \azero\ mesons.
The data sample corresponds to 89 million \BB\ pairs collected with the
\babar\ detector operating at the PEP-II asymmetric-energy $B$ Factory
at SLAC.  We find no significant signals and 
determine the following 90\% C.L. upper limits:
$\Bazerompip < \ULazerompip$, $\BazeromKp < \ULazeromKp$, 
$\BazeromKz < \ULazeromKz$, 
$\Bazerozpip < \ULazerozpip$, $\BazerozKp < \ULazerozKp$, and
$\BazerozKz < \ULazerozKz$, where in all cases $\calB$ indicates the
product of branching fractions for $B\to\azero X$ and $\azero\to\eta\pi$,
where $X$ indicates $K$ or $\pi$.

\end{abstract}


\maketitle

We report results on measurements of $B$-meson decays to charmless final states
with \azero(980) mesons \cite{az980}.  Both experimentally and theoretically, 
most work in charmless two-body $B$ decays has involved states with 
only pseudoscalar and vector mesons.  The only charmless 
$B$ decay involving scalar mesons that has been observed is $B\to f_0(980)K$ 
\cite{f0}.  There have been no previously published 
searches for $B$ decays to final states with \azero\ mesons.  
In this paper we search for the decays \azeropi, \azeroK, and \azeroKz\
for both charged and neutral \azero\ mesons.
These measurements should provide information both for $B$ decays
to scalar mesons and the nature of those mesons.

\begin{figure}[!htb]
\vspace{0.5cm}
 \includegraphics[angle=0,width=\linewidth]{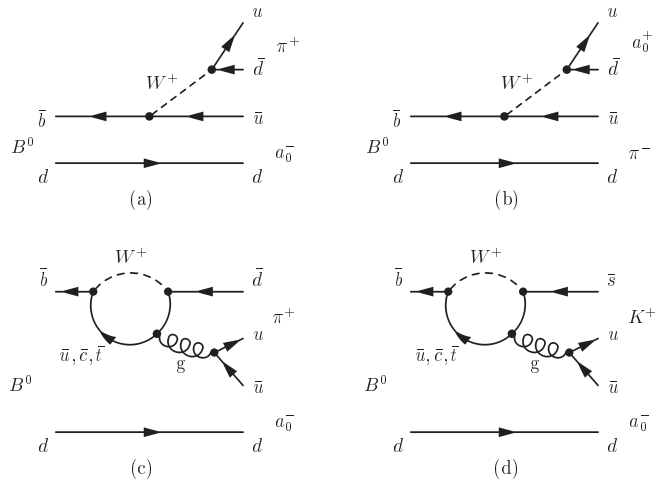}
 \caption{\label{fig:feyn}
Feynman diagrams for decays involving charged \azero\ mesons: (a) dominant and
(b) $G$-parity-suppressed tree diagrams for $\Bz\to\azero^\mp\pi^\pm$, (c) penguin
diagram for the same decay mode, and (d) penguin diagram for the decay
\azeromKp.}
\end{figure}

Some specific predictions can be made for the decays $B\to\azero\pi^\pm$ if 
factorization is assumed and if the decay is a tree or penguin (loop) process.
The dominant such process is shown in
Fig.~\ref{fig:feyn}(a).  The companion tree process, shown in
Fig.~\ref{fig:feyn}(b), is expected to be greatly suppressed, since the
virtual $W$ cannot produce an \azero\ meson \cite{vasia}.  This is a
firm prediction of the Standard Model because the weak current has a
$G$-parity even vector part and a $G$-parity odd axial-vector part. The
latter can produce an axial-vector or pseudoscalar particle
while the former produces a vector particle, but neither can
produce a $G$-parity odd scalar meson.  Penguin processes such as
shown in Fig.~\ref{fig:feyn}(c) are allowed, but are suppressed relative
to the tree processes.  Thus the
decay $B\to\azero\pi^\pm$ is expected to be ``self-tagging" (the charge of
the pion identifies the $B$ flavor).  The decays with a kaon in the final state 
should be dominated by penguin processes (Fig.~\ref{fig:feyn}d); however, 
there is a cancellation between two terms in the penguin amplitudes for these 
decays \cite{chernyak}, which leads to a prediction that the branching fraction 
should be rather small.  The diagrams for neutral $B$ decays 
involving \azeroz\ mesons are similar to those shown in Fig.~\ref{fig:feyn}.

The nature of the \azero\ is still not well understood.  It is
thought to be a \qqbar\ state with a possible admixture of a $K\Kbar$
bound-state component due to the proximity to the $K\Kbar$ threshold
\cite{PDG2002,baru}.  The \azero\ mass is known to be about 985 MeV and 
the dominant decay mode is $\azero\to\eta\pi$ \cite{PDG2002}, which is the 
mode used in the present analysis.  A recent analysis \cite{teige} that uses 
this $\eta\pi$ decay channel finds a Breit-Wigner width of ($71\pm7$) \mev,
with no better fit obtained when the more correct Flatt\' e shape 
\cite{flatte} is used.  Also since the branching fraction for 
$\azero\to\eta\pi$ is not well known, we report the
product branching fraction $\calB(B\to\azero X)\times\calB(\azero\to\eta\pi)$,
where $X$ indicates $K$ or $\pi$.

The results presented here are based on data collected
with the \babar\ detector~\cite{BABARNIM}
at the PEP-II asymmetric $e^+e^-$ collider
located at the Stanford Linear Accelerator Center.  An integrated
luminosity of 81.9~fb$^{-1}$, corresponding to 
$88.9\pm1.0$ million \BB\ pairs, was recorded at the $\Upsilon (4S)$
resonance (center-of-mass energy $\sqrt{s}=10.58\ \gev$).

The track parameters of charged particles are measured by
a combination of a silicon vertex tracker, with five
layers of double-sided silicon sensors, and a
40-layer central drift chamber, both operating in the 1.5-T magnetic
field of a superconducting solenoid. We identify photons and electrons 
using a CsI(Tl) electromagnetic calorimeter (EMC).
Further charged particle identification (PID) is provided by the average energy
loss (\dedx) in the tracking devices and by an internally-reflecting,
ring-imaging Cherenkov detector (DIRC) covering the central region.

We select $\azero$ candidates from the decay channel $\azero\to\eta\pi$ with 
the decays \etatogg\ (\etagg) and \etatoppp\ (\etappp).  We apply the following 
requirements on the invariant masses (in \mev) relevant here:
$500< m_{\gaga}<585$ for \etagg, $535<m_{\pi\pi\pi}< 560$ for
\etappp, $120 < m_{\gaga} < 150$ for \piz, and $775 < m_{\eta\pi} < 1175$ for 
$\azero\to\eta\pi$.  These requirements are typically quite loose compared
with typical resolutions in order to achieve high 
efficiency and retain sufficient sidebands to characterize the background for 
subsequent fitting.  We reconstruct \kzs\ candidates through the \kstopipi\ 
decay; to obtain a low-background, well-understood \KS\ sample, we require 
$488 < m_{\pi\pi} < 508$ \mev,
the three-dimensional flight distance from the event primary vertex to be 
greater than 2 mm, and the angle between flight and momentum vectors, in the 
plane perpendicular to the beam direction, to be less than 40 mrad.

We make several PID requirements to ensure the identity of the pions and kaons.
Secondary tracks in \etappp\ candidates must have measured DIRC, \dedx, and EMC 
outputs consistent with pions.  For the decays $B\to\azero h^+$ \cite{CC}, 
where $h^+$ indicates a charged pion or kaon, the particle $h^+$ must have an 
associated DIRC signal with a Cherenkov angle within $3.5$ standard deviations 
of the expected value for either a $\pi^\pm$ or $K^\pm$ hypothesis
(we describe below the separation between the two hypotheses).

A $B$-meson candidate is characterized kinematically by the energy-substituted 
mass $\mes=\lbrack{(\half s+\pvec_0\cdot\pvec_B)^2/E_0^2-\pvec_B^2}\rbrack^\half$
and energy difference $\DE = E_B^*-\half\sqrt{s}$, 
where $(E_B,\pvec_B)$ and $(E_0,\pvec_0)$ are the four vectors
of the $B$-candidate and the initial electron-positron system,
respectively. The asterisk denotes the \UfourS\ frame,
and $s$ is the square of the invariant mass of the electron-positron system.  
The \DE\ (\mes) resolution is about 40 MeV ($3.0\ \mev$). 
We require $|\DE|\le0.2$ GeV and $5.2\le\mes\le5.29\ \gev$.

Backgrounds arise primarily from random combinations in continuum 
$\epem\ra\qqbar$ ($q=u,d,s,c$) events. We reduce these by using the angle
\thetaT\ between the thrust axis of the $B$ candidate in the \UfourS\
frame and that of the 
rest of the charged tracks and neutral clusters in the event.
The distribution of $|\costhr|$ is
sharply peaked near $1.0$ for combinations drawn from jet-like \qqbar\
pairs, and nearly uniform for $B$-meson decays.  We require $|\costhr|<0.9$
for the $\azero\kzs$ decay modes. Based on a Monte Carlo study in which 
the relative branching fraction uncertainty is minimized, we tighten this 
requirement for the higher-background $\azero h$ 
channels: 0.8 for \fazeromhpthrpi, 0.7 for \fazeromhpgg\ and \fazerozhpthrpi,
and 0.6 for \fazerozhpgg.  We also use, in the fit 
described below, a Fisher discriminant \xf\ that combines the
angles with respect to the beam axis of the $B$ momentum and $B$ thrust axis 
(in the \UfourS\ frame), and moments describing 
the energy flow about the $B$ thrust axis \cite{PRD}.

For the \etatogg\ modes we use additional event-selection criteria to 
further reduce backgrounds from charmless $B$ decay modes such as
$B\to\Kst\gamma$ and $B\to\eta\Kst$.  We require
$|\cos{\theta^{\eta}_{\rm dec}}| \le 0.86 $, where $\theta^{\eta}_{\rm dec}$ is 
the $\eta$ decay angle, the angle of the photons in the $\eta$ rest frame with 
respect to the boost direction from the $B$ to that frame.  We also require
$\cos{\theta^{a_0}_{\rm dec}} \le 0.8$, where $\theta^{a_0}_{\rm dec}$ is the 
\azero\ decay angle, defined similarly to $\theta^{\eta}_{\rm dec}$, with 
sign such that 
high-momentum $\eta$ mesons populate the region near $+1$.  These additional
requirements reduce the \BB\ background by a factor of 2--4, depending on
the decay mode.  From Monte Carlo 
(MC) simulation \cite{geant} we estimate that the residual charmless \BB\ 
background is less than one event for all decays except \fazeromksgg\ (the 
notation indicates the decay mode of the $\eta$ used in reconstructing the 
\azero) and \fazerozhpgg, where we 
include in the fit a \BB\ component, that we find to be less than 0.5\% of
the total sample in both cases.

We obtain yields and branching fractions from extended unbinned 
maximum-likelihood fits, with input observables \DE, \mes, \xf, $m_{\eta\pi}$,
and for charged modes the PID variables $S_\pi$ and $S_K$; the last
quantities are the number of standard deviations between the measured Cherenkov
angle and the expectation for pions and kaons.

For each event $i$, hypothesis $j$ (signal, continuum background, 
\BB\ background), and, for the $\azero h^+$ decays, flavor $k$,
we define the  probability density function (PDF)
\begin{equation}
{\cal P}^i_{jk} = {\cal P}_j (\mes^i) {\cal  P}_j (\DE^i_k[,S^i_k]) { \cal P}_j(\xf^i) {\cal P}_j (m^i_{\eta\pi}).
\end{equation}
The term in brackets for $S$ pertains to the $\azero h^+$ modes.
The absence of correlations among observables (except between \DE\ and $S$,
which both depend on the momentum of the particle $h^+$)
in the background ${\cal P}^i_{jk}$, is confirmed in the
(background-dominated) data samples entering the fit.  For the signal
component, we correct for effects due to the neglect of small correlations 
(more details are provided in the systematics discussion below).  The 
likelihood function is
\begin{equation}
{\cal L} = \exp{(-\sum_{j,k} Y_{jk})}
\prod_i^{N}\left[\sum_{j,k} Y_{jk} {\cal P}^i_{jk}\right]\,,
\end{equation}
where $Y_{jk}$ is the yield of events of hypothesis $j$ and flavor $k$ that 
we find by maximizing \calL, and $N$ is the number of events in the sample.

We determine the PDF parameters from simulation for the
signal and \BB\ background components, and initial values of the
continuum background parameters from (\mes,\,\DE) sideband data.
We parameterize each of the functions ${\cal P}_{\rm sig}(\mes),\ 
{\cal  P}_{\rm sig}(\DE_k),\ { \cal P}_j(\xf),\ {\rm andi~}{ \cal P}_j(S_k)$ 
with either a Gaussian function, the sum of two Gaussian functions or an 
asymmetric Gaussian function, as required to describe the distribution.  
The component of ${\cal P}_j(m_{\eta\pi})$ which represents real \azero\
mesons in the combinatorial background is described with the same Breit-Wigner 
parameters as are used for signal.
Slowly varying distributions (\azero\ candidate mass and \DE\ 
for combinatoric background) are represented by second order Chebyshev 
polynomials.  The \qqbar\ combinatoric background in \mes\
is described by the function $f(x)=x\sqrt{1-x^2}\exp{\left[-\xi(1-x^2)\right]}$,
with $x\equiv2\mes/\sqrt{s}$ and free parameter $\xi$; for \BB\ background, 
we add a Gaussian function to the quantity $f(x)$.  Large control samples 
of $B\to D\pi$ decays of topology similar to the signal are used to verify 
the simulated resolutions in \DE\ and \mes.  Where the control data samples 
reveal differences from MC, we shift or scale the resolution used in the 
likelihood fits.  Examples of many of these PDF shapes from a very
similar analysis are shown in Ref. \cite{PRD}.
Additionally, the Breit-Wigner signal parameters for the \azero\ mass and width 
are determined from an inclusive dataset that is much larger than the 
sample used for this analysis.  The widths are consistent with
expectations from the natural-width values of Ref.~\cite{teige}.

In Table \ref{tab:results} we show for each decay mode the measured product 
branching fraction, together with the quantities entering into its 
determination.  In order to account for the uncertainties in the background
PDF descriptions, we include as free parameters in the fit, in addition to 
the signal and background yields, the principle 
parameters describing the background PDFs: slopes for the
polynomial shape for the \DE\ and \azero\ mass distributions, the
parameter $\xi$ used in the \mes\ description, and three parameters
describing the asymmetric Gaussian function for \xf.
For calculation of branching fractions, we assume that the decay rates 
of the \UfourS\ to \BpBm\ and \BzBzb\ are equal \cite{prodratio}.
We combine branching fraction results from the two $\eta$ decay
channels by adding the values of $-2\ln{\cal L}$, adjusted for a small
fit bias (see below) and taking proper account
of the correlated and uncorrelated systematic errors.

\begin{figure}[!tb]
\vspace{0.5cm}
 \includegraphics[angle=0,width=\linewidth]{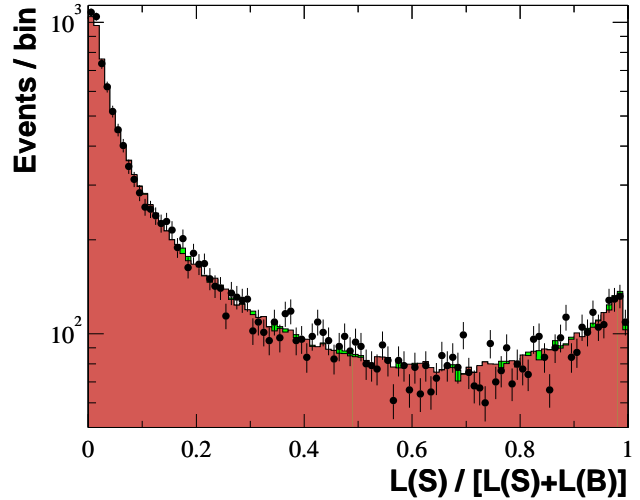}
 \caption{\label{fig:likelihood}
The likelihood ratio $\calL(S) / [\calL(S)+\calL(B)]$ for \fazeromhpgg .  The 
points represent the on-resonance data, the solid histograms are from MC
generated from background (dark shaded) and background plus signal (light 
shaded) PDFs.
  }
\end{figure}

\begin{figure}[!tb]
\vspace{0.5cm}
 \includegraphics[angle=0,width=\linewidth]{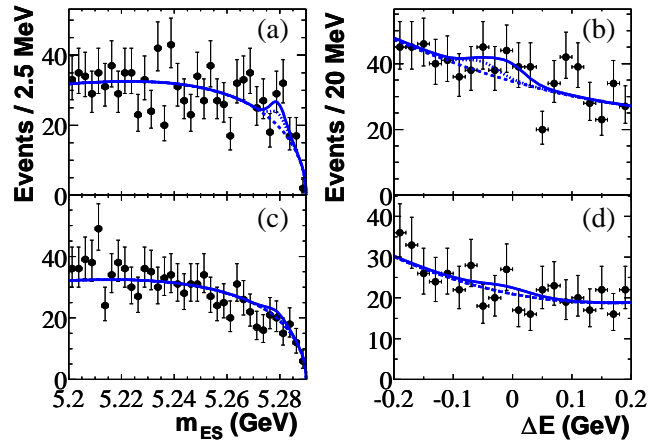}
 \caption{\label{fig:projMbDE_azeroh}
Projections of the $B$-candidate \mes\ and \DE\ for (a, b) \fazeromhp, and
(c, d) \fazerozhp.  Points with errors represent data, solid curves the full fit
functions, dashed curves the background functions (the peaking \BB\ 
background component is negligible), and the dotted curve shows the kaon
portion of the signal.  These plots are made with a minimum requirement on the 
likelihood and thus do not show all events in the data samples.
  }
\end{figure}

\begin{figure}[!htb]
\vspace{0.5cm}
 \includegraphics[angle=0,width=\linewidth]{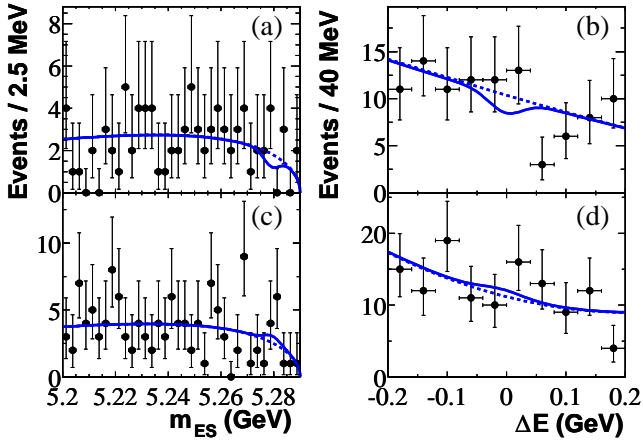}
 \caption{\label{fig:projMbDE_azeroks}
Projections of the $B$-candidate \mes\ and \DE\ for (a, b) $a_0^-\KS$, and
(c, d) $a_0^0\KS$. Points with errors represent data, 
solid curves the full fit functions, and dashed curves the background functions.
These plots are made with a minimum requirement on the likelihood and thus do 
not show all events in the data samples.
  }
\end{figure}

In order to check the suitability of the PDFs for describing the data, we show
in Fig.~\ref{fig:likelihood} the distribution of the likelihood ratio 
$\calL(S) / [\calL(S)+\calL(B)]$ for the full \fazeromhpgg\ sample, where 
$\calL(S)$ and $\calL(B)$ are the signal and background likelihood, 
respectively.  Signal would appear near one in this plot but very little is 
seen because of the small signal yield.  There is also good agreement for
similar plots for the other samples.  In order to show 
distributions of the main fit observables \mes\ and \DE, we require that
this likelihood ratio be greater than a value that would optimize the 
branching fraction uncertainty, typically 0.9 for most samples.  
In Figs.~\ref{fig:projMbDE_azeroh} and \ref{fig:projMbDE_azeroks} we show 
projections onto \mes\ and \DE\ of subsamples enriched with this requirement 
on the likelihood ratio (computed ignoring the PDF associated with the
variable plotted).

\begin{table*}[!btp]
\caption{
Signal yield, detection efficiency $\epsilon$, daughter branching fraction 
product ($\prod\calB_i$), significance (including additive systematic 
uncertainties, taken to be zero if corrected yield is negative), 
measured product branching fraction (see text), and the 90\% C.L. upper limit 
on this branching fraction.}
\label{tab:results}
\begin{tabular}{lcccccc}
\dbline
Mode	      & Yield	   &$\epsilon$(\%)&$\prod\calB_i$(\%)& Signif. &  \bfemsix & UL($10^{-6}$) \\
\tbline
~~\fazerompipgg & $18^{+11}_{-10}$&18.8&39.4&1.3&$2.3^{+1.7}_{-1.5}\pm0.9$\\
~~\fazerompipthrpi &$15^{+9}_{-8}$&15.5&22.6&1.6&$3.9^{+2.9}_{-2.5}\pm1.0$\\
\bma{\fazerompip}& & & &\bma{\sazerompip}&\bma{\razerompip}&\bma{<\ulazerompip}\\
~~\fazeromKpgg   &  $2^{+6}_{-4}$ &17.9&39.4&0.1&$0.0^{+0.9}_{-0.6}\pm0.3$\\
~~\fazeromKpthrpi &$13^{+8}_{-6}$ &14.9&22.6&1.1&$3.1^{+2.5}_{-2.1}\pm1.9$\\
\bma{\fazeromKp} & & & &\bma{\sazeromKp} &\bma{\razeromKp} &\bma{<\ulazeromKp}\\
~~\fazeromksgg   &$-12^{+8}_{-6}$ &21.4&13.5&0.0&$-3.7^{+2.9}_{-2.3}\pm 0.9$\\
~~\fazeromksthrpi & $0^{+7}_{-5}$ &15.8& 7.9&0.5& $2.7^{+6.1}_{-4.4}\pm 1.9$\\
\bma{\fazeromKz} & & & &\bma{\sazeromKz} &\bma{\razeromKz} &\bma{<\ulazeromKz}\\
~~\fazerozpipgg & $17^{+11}_{-9}$&12.8&39.4&1.4&$3.1^{+2.4}_{-2.0}\pm 1.2$\\
~~\fazerozpipthrpi &$1^{+8}_{-6}$& 9.5&22.6&0.3&$1.2^{+3.9}_{-3.2}\pm 1.7$\\
\bma{\fazerozpip}& & & &\bma{\sazerozpip}&\bma{\razerozpip}&\bma{<\ulazerozpip}\\
~~\fazerozKpgg    &$0^{+5}_{-3}$ &12.4&39.4&0.3&$0.3^{+1.1}_{-0.6}\pm 0.4$\\
~~\fazerozKpthrpi  &$6^{+7}_{-5}$& 9.1&22.6&0.5&$1.9^{+3.8}_{-2.9}\pm 2.5$\\
\bma{\fazerozKp} & & & &\bma{\sazerozKp} &\bma{\razerozKp} &\bma{<\ulazerozKp}\\
~~\fazerozksgg   & $0^{+6}_{-5}$  &15.0&13.3&0.5& $1.4^{+3.5}_{-2.4}\pm 1.2$\\
~~\fazerozksthrpi & $4^{+5}_{-4}$ & 9.7& 7.8&1.2& $6.6^{+7.8}_{-5.4}\pm 2.8$\\
\bma{\fazerozKz} & & & &\bma{\sazerozKz} &\bma{\razerozKz} &\bma{<\ulazerozKz}\\
\dbline
\end{tabular}
\vspace{-5mm}
\end{table*}

The statistical error on the signal yield is taken as the change in 
the central value when the quantity $-2\ln{\cal L}$ increases by one 
unit from its minimum value. The significance is taken as the square root 
of the difference between the value of $-2\ln{\cal L}$ (with additive systematic 
uncertainties included) for zero signal and the value at the minimum,
with other parameters free in both cases.
The 90\% confidence level (C.L.) upper limit is taken to be the branching 
fraction below which lies 90\% of the total of the likelihood integral 
(with systematic uncertainties included)
in the positive branching fraction region.

Most of the yield uncertainties arising from lack of knowledge of the PDFs have
been included in the statistical error since most background parameters are 
free in the fit.  Varying the signal PDF parameters within their estimated 
uncertainties, we determine the uncertainties in the signal PDFs to be 1--5 
events, depending on the final state.  The contribution to this uncertainty 
from the parameterization of the \azero\ signal shape is small.  We verify
that the value of the likelihood of each fit is consistent with the expectation
found from an ensemble of simulated experiments.

Uncertainties in our knowledge of the efficiency, found from auxiliary 
studies, include 0.8\%$\cdot N_t$, 2.5\%$\cdot N_\gamma$, and 4\%\ for a
\KS\ decay, where $N_t$ and $N_\gamma$ are the number of signal tracks
and photons, respectively.  Our estimate of the 
number of produced \BB\ events is uncertain by 1.1\%.  
The neglect of correlations among observables in the fit can cause a systematic 
bias; the correction for this bias (between $-$3 and $+3$ events) and assignment
of the resulting systematic uncertainty (0.5--2 events) is determined from 
simulated samples with varying background populations.  Published data
\cite{PDG2002}\ provide the uncertainties in the $B$-daughter product branching 
fractions (1--2\%).  Selection efficiency uncertainties are 0.5--3.5\% for
\costhr\ and 0.5\% for PID (for the $\azero h^+$ modes).

In conclusion, we do not find significant signals for these $B$-meson decays to
states with \azero\ mesons.  The measured branching fractions and 90\% C.L. 
upper limits are given in Table \ref{tab:results}.  
Assuming $\eta\pi$ to be the dominant \azero\ decay mode, we rule out
the predictions for the decay \azeromKz\ derived in Ref. \cite{minkochs}.

We are grateful for the excellent luminosity and machine conditions
provided by our \pep2\ colleagues, 
and for the substantial dedicated effort from
the computing organizations that support \babar.
The collaborating institutions wish to thank 
SLAC for its support and kind hospitality. 
This work is supported by
DOE
and NSF (USA),
NSERC (Canada),
IHEP (China),
CEA and
CNRS-IN2P3
(France),
BMBF and DFG
(Germany),
INFN (Italy),
FOM (The Netherlands),
NFR (Norway),
MIST (Russia), and
PPARC (United Kingdom). 
Individuals have received support from CONACyT (Mexico), A.~P.~Sloan Foundation, 
Research Corporation,
and Alexander von Humboldt Foundation.


\begin{thebibliography}{99}

\bibitem{az980}
Throughout this note, when we refer to \azero, we mean specifically \azero(980).

\bibitem{f0}
Belle Collaboration, A. Garmash \etal, \jprd{65}, 092005 (2002);
\babar\ Collaboration, B. Aubert \etal, hep-ex/0308065 (submitted to \prd{}), 2003;
\babar\ Collaboration, B. Aubert \etal, hep-ex/0406040 (submitted to \jprl{}), 2004.

\bibitem{CC}
The named member of a charge-conjugate pair of particles stands for either.

\bibitem{vasia}
S. Laplace and V. Shelkov, \epjc{22}, 431 (2001).

\bibitem{chernyak}
V. Chernyak, \plb{509}, 273 (2001).

\bibitem{PDG2002}
Particle Data Group, K.~Hagiwara \etal, \jprd{66}, 010001 (2002).

\bibitem{baru}
V. Baru \etal, \plb{586}, 53 (2004).

\bibitem{teige}
S. Teige \etal, \jprd{59}, 012001 (1998).

\bibitem{flatte}
S. Flatt\' e, \plb{63}, 224 (1976).

\bibitem{BABARNIM}
\babar\ Collaboration, B.\ Aubert \etal, \nima{479}, 1 (2002).

\bibitem{PRD}
\babar\ Collaboration, B.\ Aubert \etal, \jprd{70}, 032006 (2004).

\bibitem{geant}
The \babar\ detector Monte Carlo simulation is based on GEANT4:
S. Agostinelli \etal, \nima{506}, 250 (2003).

\bibitem{prodratio}
See for instance \babar\ Collaboration, B.\ Aubert \etal, \jprd{69}, 071101 (2004) and references therein.

\bibitem{minkochs}
P. Minkowski and W. Ochs, hep-ph/0404194 (2004).

\end{thebibliography}
\end{document}